\documentclass{aa}

\usepackage{graphicx}
\usepackage{txfonts}
\usepackage{natbib}     
\usepackage{xcolor} 
\usepackage{threeparttable} 
\usepackage[colorlinks=true,citecolor=blue, linkcolor=blue]{hyperref} 
\usepackage{float} 
\usepackage{multirow}

\bibpunct{(}{)}{;}{a}{}{,} 

\begin{document}

        \title{Detection of iron emission lines and a temperature inversion \\ on the dayside of the ultra-hot Jupiter KELT-20b}

   \author{F. Yan\inst{1}
        \and
                A. Reiners\inst{1}
          \and
          E. Pall\'e\inst{2,3}
          \and
          D. Shulyak\inst{4,5}
          \and
                        M.~Stangret\inst{2,3}
                        \and
                        K. Molaverdikhani\inst{6,7,8}
          \and
          L.~Nortmann\inst{1}
          \and
          P.~Molli\`ere\inst{7}
         \and
          Th.~Henning\inst{7}
          \and
                N.~Casasayas-Barris\inst{9}
                \and
                D. Cont\inst{1}
          \and
          G.~Chen\inst{10,11}
          \and
          S.~Czesla\inst{12,13}
          \and
           A. S\'anchez-L\'opez\inst{9}
          \and
          M. L\'opez-Puertas\inst{5}          
          \and
                I.~Ribas\inst{14,15},  A.~Quirrenbach\inst{6}, J.~A.~Caballero\inst{16}, P.~J.~Amado\inst{5}
                \and
            D.~Galad\'i-Enr\'iquez\inst{17}, S.~Khalafinejad\inst{6}, L.~M.~Lara\inst{5}, D.~Montes\inst{18}, G.~Morello\inst{2}, E.~Nagel\inst{12,13}, E.~Sedaghati\inst{5}, M. R. Zapatero Osorio\inst{19}
            \and M.~Zechmeister\inst{1}
             }
                
\institute{Institut f\"ur Astrophysik, Georg-August-Universit\"at, Friedrich-Hund-Platz 1, 37077 G\"ottingen, Germany\\
        \email{fei.yan@uni-goettingen.de}
\and    
        Instituto de Astrof{\'i}sica de Canarias (IAC), V{\'i}a Lactea s/n, 38200 La Laguna, Tenerife, Spain
\and
Departamento de Astrof{\'i}sica, Universidad de La Laguna, 38026  La Laguna, Tenerife, Spain
\and
Max-Planck-Institute f\"ur Sonnensystemforschung, Justus-von-Liebig-Weg 3, 37075 G\"ottingen, Germany
\and
Instituto de Astrof\'{\i}sica de Andaluc\'{\i}a - CSIC, Glorieta de la Astronom\'{\i}a s/n, 18008 Granada, Spain
\and
Landessternwarte, Zentrum f\"ur Astronomie der Universit\"at Heidelberg, K\"onigstuhl 12, 69117 Heidelberg, Germany
\and
Max-Planck-Institut f{\"u}r Astronomie, K{\"o}nigstuhl 17, 69117 Heidelberg, Germany
\and
Ludwig-Maximilians-Universit\"at, Universitäts-Sternwarte München, Scheinerstr. 1, 81679, Munich, Germany
\and
        Leiden Observatory, Leiden University, Postbus 9513, 2300 RA, Leiden, The Netherlands
\and
CAS Key Laboratory of Planetary Sciences, Purple Mountain Observatory, Chinese Academy of Sciences, Nanjing 210023, China
\and    
CAS Center for Excellence in Comparative Planetology, Hefei 230026, China
\and
Hamburger Sternwarte, Universit{\"a}t Hamburg, Gojenbergsweg 112, 21029 Hamburg, Germany
\and
Th{\"u}ringer Landessternwarte Tautenburg, Sternwarte 5, 07778 Tautenburg, Germany
\and
Institut de Ci\`encies de l'Espai (CSIC-IEEC), Campus UAB, c/ de Can Magrans s/n, 08193 Bellaterra, Barcelona, Spain
\and
Institut d'Estudis Espacials de Catalunya (IEEC), 08034 Barcelona, Spain
\and
 Centro de Astrobiolog\'{i}a (CSIC-INTA), ESAC, Camino bajo del castillo s/n, 28692 Villanueva de la Ca\~nada, Madrid, Spain
\and
Centro Astron{\'o}nomico Hispano-Alem{\'a}n (CSIC--Junta de Andaluc\'ia), Observatorio Astron{\'o}nomico de Calar Alto, Sierra de los Filabres, 04550 G{\'e}rgal, Almer\'ia, Spain
\and
Departamento de F\'{i}sica de la Tierra y Astrof\'{i}sica 
and IPARCOS-UCM (Instituto de F\'{i}sica de Part\'{i}culas y del Cosmos de la UCM), 
Facultad de Ciencias F\'{i}sicas, Universidad Complutense de Madrid, 28040, Madrid, Spain
\and
Centro de Astrobiolog\'{i}a (CSIC-INTA), Carretera de Ajalvir, km 4, 28850 Torrej\'{o}n de Ardoz, Madrid, Spain
\\      }
        \date{Received 8 October 2021; accepted 15 January 2022}


  \abstract
{Ultra-hot Jupiters (UHJs) are gas giants with very high equilibrium temperatures. In recent years, multiple chemical species, including various atoms and ions, have been discovered in their atmospheres. Most of these observations have been performed with transmission spectroscopy, although UHJs are also ideal targets for emission spectroscopy due to their strong thermal radiation. We present high-resolution thermal emission spectroscopy of the transiting UHJ KELT-20b/MASCARA-2b. The observation was performed with the CARMENES spectrograph at orbital phases before and after the secondary eclipse. We detected atomic Fe using the cross-correlation technique. The detected Fe lines are in emission, which unambiguously indicates a temperature inversion on the dayside hemisphere. We furthermore retrieved the temperature structure with the detected Fe lines. The result shows that the atmosphere has a strong temperature inversion with a temperature of $4900\pm{700}$ K and a pressure of $10^{-4.8_{-1.1}^{+1.0}}$ bar at the upper layer of the inversion. A joint retrieval of the CARMENES data and the TESS secondary eclipse data returns a temperature of $2550_{-250}^{+150}$ K and a pressure of $10^{-1.5_{-0.6}^{+0.7}}$ bar at the lower layer of the temperature inversion. 
The detection of such a strong temperature inversion is consistent with theoretical simulations that predict an inversion layer on the dayside of UHJs. 
The joint retrieval of the CARMENES and TESS data demonstrates the power of combing high-resolution emission spectroscopy with secondary eclipse photometry in characterizing atmospheric temperature structures.
}

   \maketitle

%
~

\section{Introduction}
Ultra-hot Jupiters (UHJs) are giant exoplanets with very high dayside temperatures that typically exceed 2000\,K. Theoretical studies suggest that their properties are different from those of planets with more modest temperatures. For example, their dayside atmospheres can be dominated by atoms and ions with a large number of molecules that are thermally dissociated \citep[e.g.,][]{Lothringer2018, Parmentier2018, Kitzmann2018, Fossati2020}. These planets have extreme differences in their dayside to nightside temperature  and are also different chemically  \citep[e.g.,][]{Bell2018, Komacek2018, Tan2019, Helling2019, Molaverdikhani2020}. Simulations also indicate that the dayside hemispheres of these planets have temperature inversion layers because the absorption of the stellar radiation by species such as metals and metal oxides is strong \citep[e.g.,][]{Lothringer2019,Gandhi2019-inversion, Baxter2020}. It has recently been shown that for the hottest exoplanet KELT-9b, the deviation from local thermodynamic equilibrium in the level population of \ion{Fe}{ii} is the main driver of strong temperature inversion in the high-altitude atmosphere \citep{Fossati2021-NLTE}.

Various chemical species have been detected in UHJs via transmission spectroscopy. For example, hydrogen Balmer lines and various metals (including \ion{Fe}{i}, \ion{Fe}{ii}, \ion{Ti}{i}, \ion{Ca}{ii}, \ion{Mg}{i}) have been detected in the transmission spectrum of KELT-9b \citep{Yan2018, Hoeijmakers2018, Hoeijmakers2019, Yan2019, Cauley2019, Turner2020, Wyttenbach2020}. Hydrogen Balmer lines and \ion{Ca}{ii} have also been found in the atmosphere of WASP-33b \citep{Yan2019,Yan2021-W33,Cauley2021-W33, Borsa2021-W33}. In the transmission spectrum of WASP-76b, several metals including \ion{Fe}{i}, \ion{Na}{i}, \ion{Ca}{ii,} and \ion{Li}{i} have been discovered \citep{Seidel2019, Zak2019, Ehrenreich2020, Tabernero2021, Casasayas-Barris2021}. Hydrogen Balmer lines and various metals have also been detected in the inflated atmosphere of WASP-121b \citep{Sing2019, Bourrier2020, Gibson2020,Cabot2020,Ben-Yami2020, Borsa2021-W121}.

Their ultra-high dayside temperatures also make UHJs  ideal targets for thermal emission observations. For example, near-infrared emission spectra have been observed with the Hubble Space Telescope (HST) for WASP-12b \citep{Stevenson2014}, Kepler-13Ab \citep{Beatty2017}, WASP-18b \citep{Arcangeli2018}, HAT-P-7b \citep{Mansfield2018}, WASP-103b \citep{Kreidberg2018}, WASP-76b \citep{Edwards2020}, KELT-7b \citep{Pluriel2020}, and KELT-9b \citep{Changeat2021}. Secondary eclipses and phase curves of several UHJs have also been observed at the optical wavelengths with Kepler, TESS, and CHEOPS \citep[e.g.,][]{Zhang2018, Essen2020,Bourrier2020-TESS, Wong2020,Mansfield2020,Lendl2020, Daylan2021}. 
Thermal emission spectroscopy is particularly sensitive to the temperature structure of the dayside hemisphere, and it has therefore been used to probe the temperature inversion layers in UHJs. For example, \cite{Evans2017} detected temperature inversion in WASP-121b by observing the $\mathrm{H_2O}$ emission band with the HST.
Evidence of temperature inversions has also been inferred in several UHJs from measurement of the infrared CO emission feature with the Spitzer telescope \citep[e.g.,][]{Sheppard2017, Kreidberg2018}.

Recently, atomic iron has been detected in the high-resolution thermal emission spectra of three UHJs WASP-189b \citep{Yan2020}, KELT-9b \citep{Pino2020,Kasper2021}, and  WASP-33b \citep{Nugroho2020W33, Cont2021}. In addition to Fe, other species such as OH and TiO emission lines have been detected in WASP-33b \citep{Nugroho2017, Nugroho2021,Herman2020, Cont2021}.
The detected spectral lines of these chemical species are all in emission, which means that the flux of the spectral line is higher than that of the continuum.
In a thermal radiation spectrum, the flux of the spectral line originates from a higher altitude than the adjacent continuum.
Thus, an emission line profile indicates a hotter temperature at a higher altitude.
Therefore, the detected emission spectral features in the three UHJs are unambiguous evidence for temperature inversion layers in the dayside atmosphere of UHJs.


Thermal structure is a key property of planetary atmospheres. The existence and origin of temperature inversions has long been an open question in the field of exoplanets. Temperature inversions in hot Jupiters were initially proposed by \cite{Hubeny2003} and \cite{Fortney2008}, who suggested that the strong absorption of titanium oxide (TiO) and vanadium oxide (VO) can create an inversion. Theoretical simulations later suggested that atomic metals such as Fe are also capable of producing temperature inversions in UHJs \citep[e.g.,][]{Arcangeli2018,Lothringer2018}.
Therefore, the recent detection of emission lines (e.g., Fe and OH) using high-resolution spectroscopy is an important advance in understanding the presence and origin of temperature inversions.


Here, we report the detection of \ion{Fe}{i} emission lines in the dayside spectrum of \object{KELT-20b}/\object{MASCARA-2b}. The planet is an ultra-hot Jupiter with an equilibrium temperature ($T_\mathrm{eq}$) of $\sim$ 2300\,K. The transmission spectrum of the planet has been observed with several different instruments and various spectral features have been detected, including hydrogen Balmer lines, \ion{Fe}{i}, \ion{Fe}{ii}, \ion{Na}{i}, \ion{Ca}{ii}, \ion{Mg}{i}, and \ion{Cr}{ii} \citep{Casasayas-Barris2018, Casasayas-Barris2019, Stangret2020, Nugroho2020-KELT20, Hoeijmakers2020, Kesseli2020, Rainer2021}. We present the first thermal emission spectroscopy of this planet.

The paper is organized as follows. In Section 2 we describe the observations and data reduction. In Section 3 we present the results and discussions on the detection of \ion{Fe}{i} emission lines and the retrieval of atmospheric structure. The conclusion is presented in Section 4.

%
\section{Observations and data reduction}
\begin{figure}
   \centering
   \includegraphics[width=0.50\textwidth]{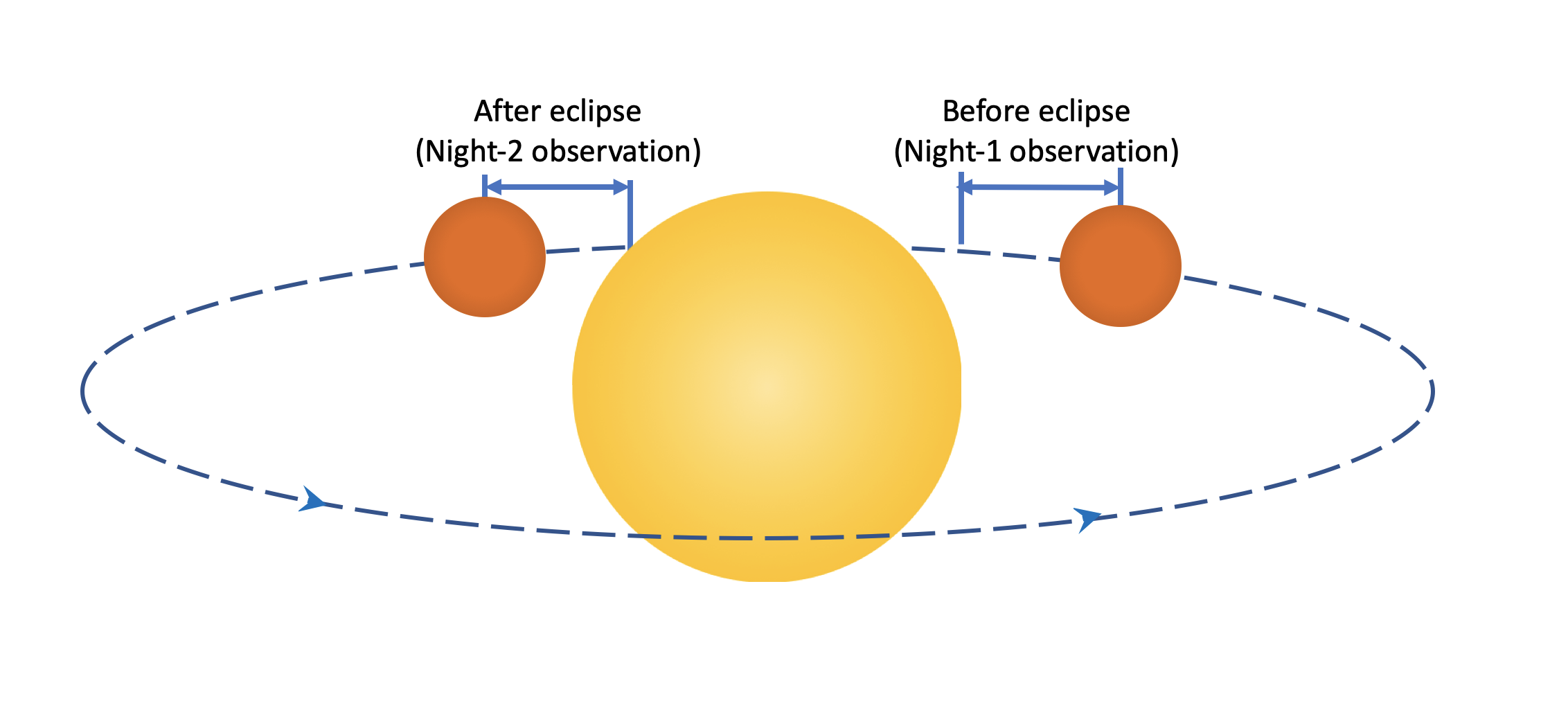}
      \caption{Schematic of the thermal emission observations. The orbital phase coverage of the two nights of observations is also indicated. The figure is not to scale.
      }
         \label{fig-obs}
   \end{figure}

We observed the thermal emission spectrum of \object{KELT-20b} with the CARMENES spectrograph \citep{Quirrenbach2018}, installed at the 3.5 m telescope of the Calar Alto Observatory, on 21 May 2020 and 9 July 2020. The visual channel of the spectrograph has a high spectral resolution (\textit{R} $\sim$ 94\,600) and a wavelength coverage of 520--960\,nm. The wavelength coverage of the visual channel is continuous without a gap between the spectral orders. The instrument has two fibers with a size of 1.5\,arcsec projected on the sky. We located fiber A on the target and fiber B on the background sky (at 88\,arcsec to the east of the target).
The observing time was carefully chosen so that the first night of observations covered the orbital phases before the secondary eclipse, and the second night of observations covered the phases after eclipse (Fig. \ref{fig-obs}). 
The observation in the first night was continuous, but that of the second night  was interrupted by an instrumental software issue for about 10 minutes. We also discarded two spectra from the second night because their flux levels were low. In total, we obtained 168 spectra, 13 of which were taken during the secondary eclipse.
The detailed observation logs are summarized in Table \ref{obs_log}.

The raw spectra were reduced with the CARMENES pipeline {\tt CARACAL v2.20} \citep{Zechmeister2014, Caballero2016}. 
The pipeline produces wavelength solutions that are obtained from the calibration lamps. These wavelength solutions are normally precise enough for detecting exoplanet atmospheres. For example, \cite{Alonso-Floriano2019} found that the CARMENES wavelengths drift during the night is about 15\,m/s, which is negligible for exoplanet atmosphere observations. Therefore, we did not apply any further wavelength correction.
The pipeline provides one-dimensional spectra with 61 spectral orders in the observatory rest frame along with a noise estimate for each data point. The detailed noise estimation of the pipeline was described by \cite{Zechmeister2014}. We used the original wavelength sampling provided by the pipeline.
We normalized the spectra order by order using a seventh-order polynomial fit.

\begin{figure}
   \centering
   \includegraphics[width=0.49\textwidth]{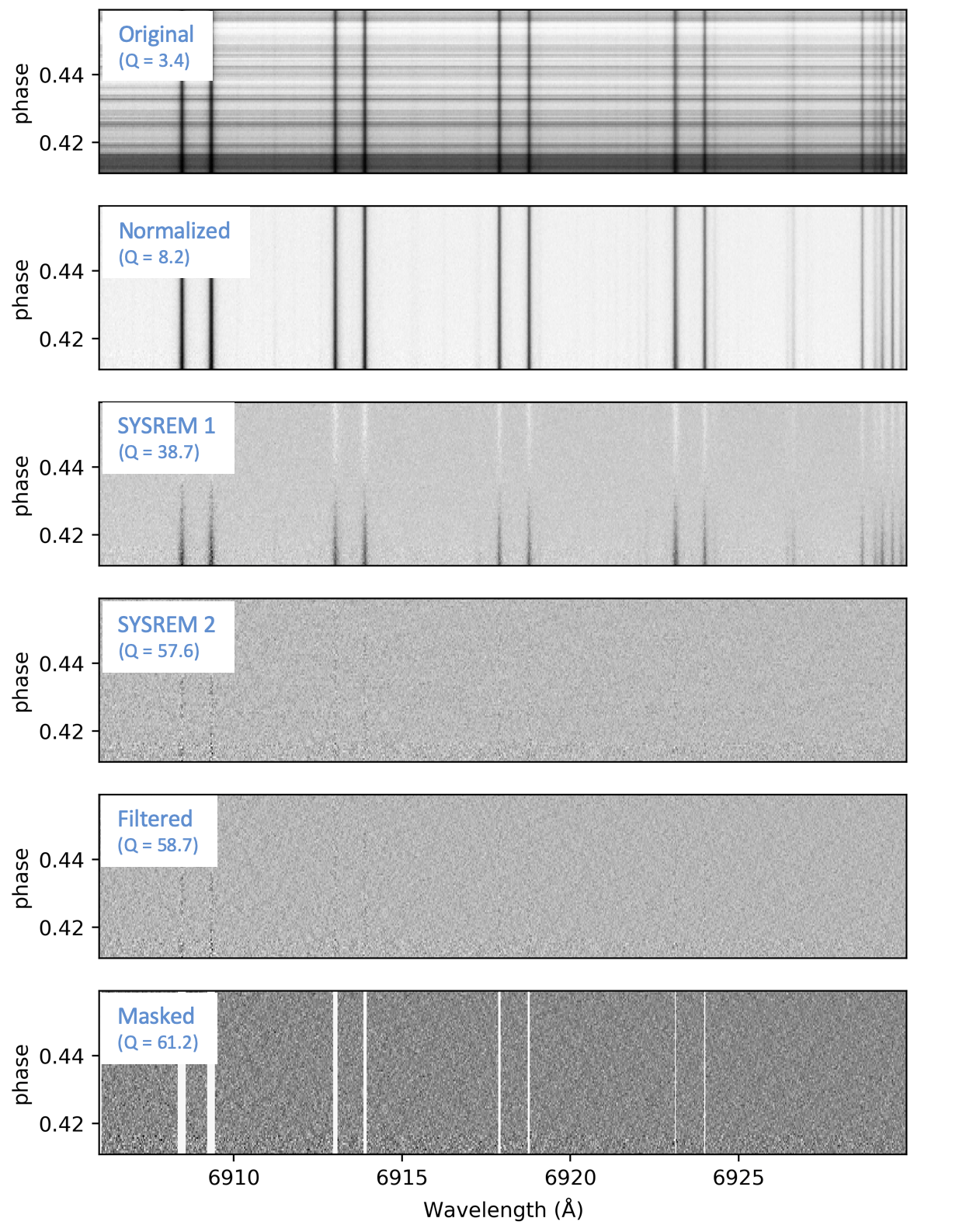}
      \caption{Example of the data reduction procedure. These are the spectra from the first observation night, which are presented in a small wavelength range for demonstration purposes. The spectral matrices from top to bottom are original spectra, normalized spectra, spectra after the first \texttt{SYSREM}, spectra after the second \texttt{SYSREM}, spectra after the Gaussian high-pass filtering, and spectra after the masking. To estimate the efficiency of each data reduction procedure, we computed a metric $Q$, which is defined as the ratio of the mean value and the standard deviation of each spectral matrix. 
      }
         \label{Demo}
   \end{figure}

\begin{figure}
   \centering
   \includegraphics[width=0.45\textwidth]{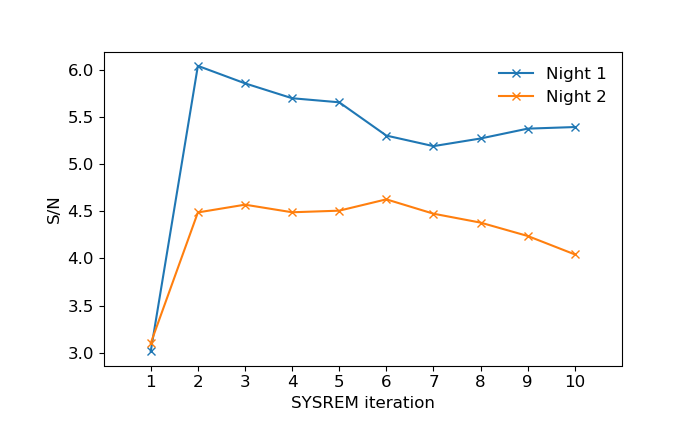}
      \caption{Evolution of the detection significance (i.e., S/N at the best-fit $K_\mathrm{p}$-$\mathrm{\Delta} \varv$ location) with different \texttt{SYSREM} iteration numbers. 
      }
         \label{APP-sysrem}
   \end{figure}

We removed the telluric and stellar lines using the \texttt{SYSREM} algorithm \citep{Tamuz2005, Birkby2013}. The input data for \texttt{SYSREM} are the normalized order-by-order spectral matrix. The \texttt{SYSREM} algorithm also requires the noise of each data point. We used the noise value from the pipeline and applied error propagation to obtain the noise of the normalized spectrum. 
To preserve the relative depths of the planetary spectral lines, we applied the method proposed by \cite{Gibson2020}. We first performed the \texttt{SYSREM} iterations in flux-space and instead of subtracting the input data with the \texttt{SYSREM} model, we then divided the input data by the \texttt{SYSREM} model. With this procedure, we preserved the strength of the planetary spectral lines at locations of the stellar and telluric absorption lines. The \texttt{SYSREM} procedure was performed on each spectral order separately.
We tested different \texttt{SYSREM} iteration numbers (1 to 10) for the data set of the two nights.  The final results do not change significantly after the first two iterations (cf. Fig.~\ref{APP-sysrem}). We chose to use two iterations for night 1 and six iterations for night 2 because the final detection significance is highest at these iterations.
The output spectra from \texttt{SYSREM} were then shifted into the stellar rest frame by correcting for the stellar systemic velocity \citep[--24.48 km\,s$^{-1}$,][]{Rainer2021} and the observer's barycentric radial velocity (RV) of Earth. 
To further remove any broadband features in the residual spectrum, we filtered the spectra using a Gaussian high-pass filter with a Gaussian $\sigma$ of 15 points ($\sim$ 19 km\,s$^{-1}$). 
These final residual spectra were used to search for planetary spectral lines.
An example of the data reduction procedure is presented in Fig. \ref{Demo}.

%
\begin{table*}
\caption{Observation logs.}             
\label{obs_log}      
\centering                          
\begin{threeparttable}
        \begin{tabular}{l c c c c c c c}        
        \hline\hline \noalign{\smallskip}                
          &  Date &  Airmass change & Exposure time [s] & $N_\mathrm{spectra}$ & Phase coverage & $S/N$ range \tablefootmark{a} \\     
        \hline \noalign{\smallskip}                      
 Night-1 & 2020-05-21     & 1.87--1.01 & 120 & 85 & 0.411--0.459 & 52-107\\ 
Night-2 & 2020-07-09      & 1.07--1.01--1.17 & 120 & 83 & 0.515--0.564 & 65-92\\                         
\hline                                 
        \end{tabular}
        \tablefoot{
\tablefoottext{a}{The $S/N$ per pixel was measured at $\sim$ 6510\,$\mathrm{\AA}$. }
}
\end{threeparttable}      
\end{table*}

%
\begin{table*}
\caption{Parameters of system KELT-20b/MASCARA-2b.}             
\label{paras_K20}      
\centering   
\begin{threeparttable}                       
\begin{tabular}{l c c  }        
\hline\hline  \noalign{\smallskip}               
        Parameter & Symbol [unit] & Value   \\
        \hline  \noalign{\smallskip}                     
        \textit{The star} & ~ & ~ \\
                Effective temperature           & $T_\mathrm{eff} [K]$  &       8980$_{-130}^{+90}$ \tnote{a}   \\
        Radius  & $R_\star$ [$R_\odot$] &       1.60 $\pm$ 0.06 \tnote{a}    \\
        Mass    & $M_\star$     [$M_\odot$]&    1.89$_{-0.05}^{+0.06}$ \tnote{a}    \\
                Systemic velocity & $v_\mathrm{sys}$ [km\,s$^{-1}$]     &       --24.48 $\pm$ 0.04 \tnote{b}   \\                       
        ~ & ~ & ~  \\
        \textit{The planet} & ~ & ~  \\
        Radius          & $R_\mathrm{p}$ [$R_\mathrm{J}$]       & 1.83 $\pm$ 0.07 \tnote{a}    \\
        Mass    & $M_\mathrm{p}$        [$M_\mathrm{J}$]        &       < 3.51 \tnote{c}    \\
        Surface gravity & log\,$g$      [log cgs]       &       < 3.42 \tnote{c}    \\
        Orbital period & $P$ [d]        &       $3.4741070\pm{0.0000019}$ \tnote{c}  \\
        Transit epoch (BJD) & $T_\mathrm {0}$ [d]       &        $2457503.12005\pm{0.00019}$ \tnote{c}  \\
        Transit duration & $T_\mathrm {14}$ [d] & $0.14898\pm{0.0009}$ \tnote{c} \\
                RV semi-amplitude & $K_\mathrm{p}$ [km\,s$^{-1}$] & 175.5$_{-2.3}^{+2.8}$ \tnote{a}    \\ \noalign{\smallskip}
                ~ & ~ & 169.3$_{-6.9}^{+8.9}$ \tnote{c} \\ \noalign{\smallskip}

\hline                               
\end{tabular}
\tablefoot{
  \tablefoottext{a}{\cite{Talens2018}.}
  \tablefoottext{b}{\cite{Rainer2021}.}
  \tablefoottext{c}{\cite{Lund2017}. There are two different $K_\mathrm{p}$ values because the stellar masses in the corresponding publications are different. }
}
\end{threeparttable}      
\end{table*}

%

\section{Results and discussions}
\subsection{Detection of Fe emission lines}
Because \ion{Fe}{i} has relatively strong and dense emission lines in the CARMENES wavelength range, it is an ideal chemical species for emission spectroscopy observations.
To search for the \ion{Fe}{i} lines in the thermal emission spectrum, we cross-correlated the observed spectrum with a theoretical template spectrum \citep{Snellen2010}.

\subsubsection{Cross-correlation method}
\begin{figure}
   \centering
   \includegraphics[width=0.45\textwidth]{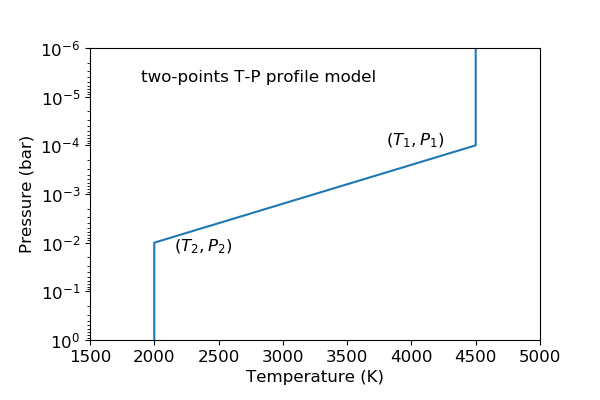}
   \includegraphics[width=0.49\textwidth]{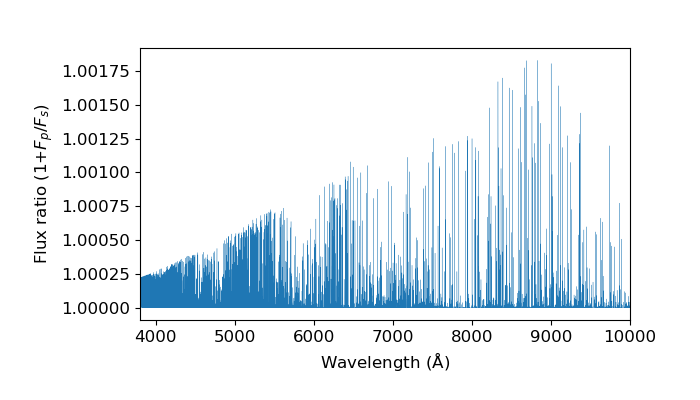}
      \caption{\textit{} Two-point $T$-$P$ profile for the template spectrum calculation (top). Calculated thermal emission spectrum of \ion{Fe}{i} (bottom). This template spectrum was used for the cross-correlation. 
      }
         \label{template}
   \end{figure}

We calculated the template spectrum similarly as described in \cite{Yan2020}. We assumed a two-point parameterized temperature-pressure ($T$-$P$) profile \citep{Brogi2014}. At altitudes above the lower pressure point ($T_\mathrm{1}$, $P_\mathrm{1}$) and below the higher pressure point ($T_\mathrm{2}$, $P_\mathrm{2}$), the atmosphere was assumed to be isothermal, while between the two points, the temperature was assumed to change linearly with log\,$P$. According to theoretical simulations by \cite{Lothringer2019}, the two-point model is analogous to the temperature profiles of UHJs around hot stars.
For the case of KELT-20b, we set the two points as (4500\,K, $10^{-4}$\,bar) and (2000\,K, $10^{-2}$\,bar) (Fig.\ref{template}). This $T$-$P$ profile has a strong temperature inversion and is a reasonable approximation to the $T$-$P$ profiles from theoretical simulations \citep[e.g.,][]{Lothringer2019}.
We also assumed solar metallicity and set a constant mixing ratio of \ion{Fe}{i} ($10^{-4.59}$). Because the mass of the planet is not well determined and only an upper limit has been reported in the discovery paper (Table \ref{template}), we assumed a surface gravity (log\,$g$) of 3.0 for the template calculation. We then used the \texttt{petitRADTRANS} tool \citep{Molliere2019} to calculate the thermal emission spectrum of the planet ($F_\text{p}$) and obtained the stellar spectrum ($F_\text{s}$, assumed to be a blackbody spectrum). The high-resolution mode of \texttt{petitRADTRANS} provides spectra with a resolution of $10^6$.
The observed spectrum of the star and planet system should be $F_\text{p}$+$F_\text{s}$. However, because the final observed spectrum is normalized, we expressed the template spectrum as 1 + $F_\text{p}$/$F_\text{s}$. We then normalized this template spectrum by dividing it with the continuum to remove the planetary continuum spectrum.
The template was further convolved with the instrumental profile using the \texttt{broadGaussFast} code from \texttt{PyAstronomy} \citep{Czesla2019}. Here we used a Gaussian profile corresponding to the instrumental resolution of 94\,600, which is measured from the calibration lamps by the CARMENES consortium team.
Fig. \ref{template} presents the final normalized and convolved template spectrum. We subsequently generated a grid of the template spectrum by shifting the spectrum from --\,500 km\,s$^{-1}$ to +\,500 km\,s$^{-1}$ in 1 km\,s$^{-1}$ steps with a linear interpolation. The actual wavelength sampling of the instrument ranges from 1.0 km\,s$^{-1}$ to 1.5 km\,s$^{-1}$, so the 1 km\,s$^{-1}$ step is a good approximation with a slight oversampling.
We also filtered the template spectra with a Gaussian high-pass filter in the same way as was applied to the observed data, although this filtering process modifies the model spectra only slightly.

Before performing the cross-correlation, we masked the wavelength points with a low signal-to-noise ratio (S/N). We calculated an average S/N spectrum of all the observed spectra for each night and masked the wavelength points with S/N < 50 for night 1 and S/N < 40 for night 2. In this way, the strong telluric absorption lines were masked out. In addition, we excluded the data points at wavelengths below 545\,nm and above 892\,nm, considering that the spectrum has low S/Ns or strong telluric absorption lines at these wavelengths.
We then computed the cross-correlation function (CCF) of each residual spectrum by cross-correlating the spectrum with the template grid.

\subsubsection{Results of the cross-correlation}
\begin{figure}
   \centering
   \includegraphics[width=0.5\textwidth]{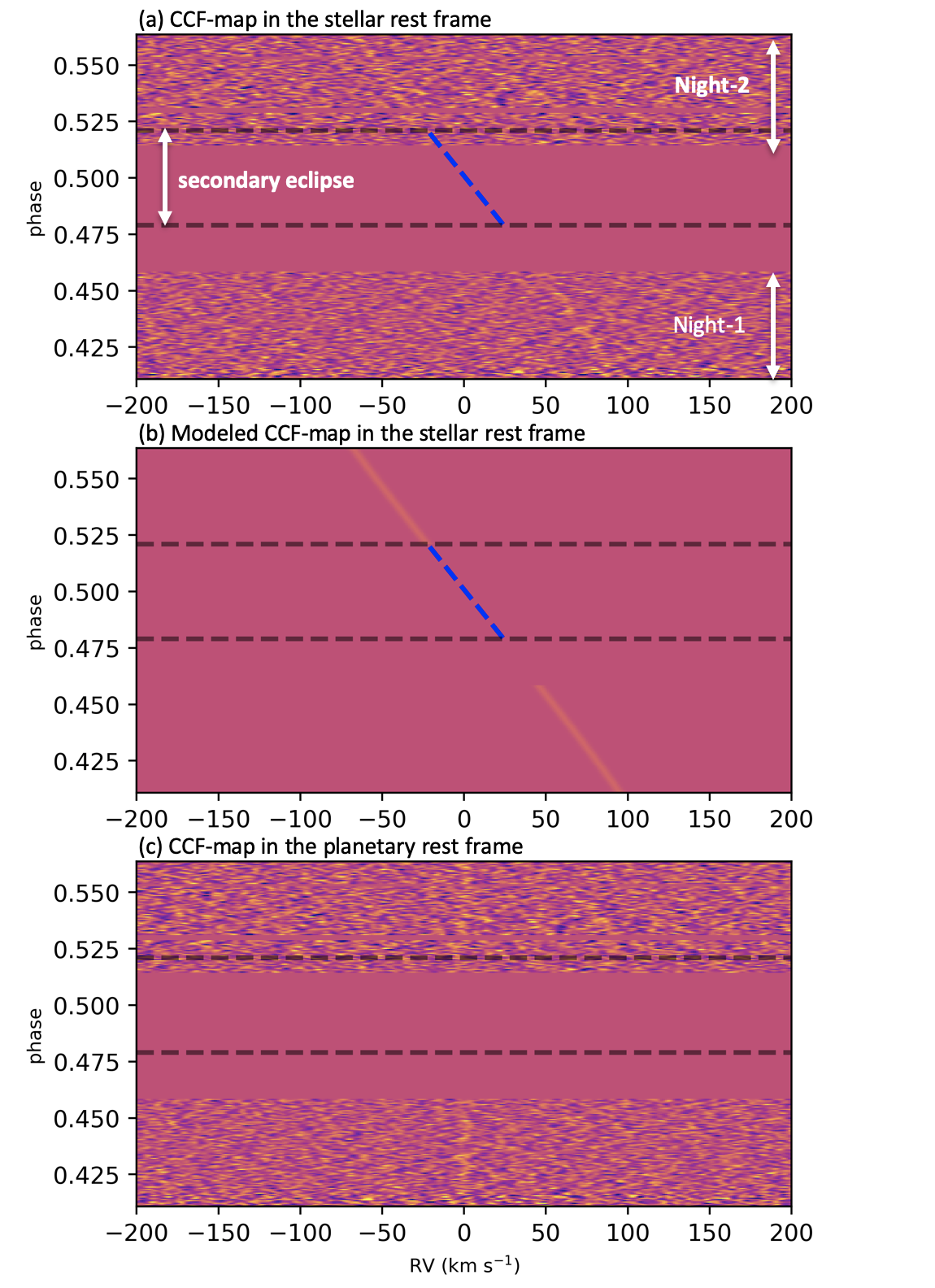}
      \caption{Cross-correlation functions of the spectra from the two nights. \textit{Upper panel:} CCF map in the stellar rest frame. \textit{Middle panel:} Modeled CCF map with the best-fit $K_\mathrm{p}$ and $\mathrm{\Delta} \varv$ values.
      \textit{Lower panel:} CCF map in the planetary rest frame.
      The horizontal dashed lines indicate the beginning and end of the eclipse. The dashed blue line denotes the planetary orbital motion RV. 
      }
         \label{CCF-map}
   \end{figure}

The obtained CCFs of the two nights spectra are presented in Fig.\ref{CCF-map}. The upper panel of the figure is the CCF-map in the stellar rest frame and the atmospheric signature is the bright stripe with positive RV before eclipse and negative RV after eclipse, reflecting the planetary orbital motion. We further modeled the CCF map using the same method as \cite{Yan2020}. We assumed that the CCF has a Gaussian profile and that the peak of the CCF is located at $K_\mathrm{p} \mathrm{sin(2\pi}\phi) + \mathrm{\Delta} \varv$, where $K_\mathrm{p}$ is the semi-amplitude of the planetary orbital RV, $\mathrm{\Delta} \varv$ is the RV deviation from the orbital motion, and $\phi$ is the orbital phase (phase 0 represents mid-transit). We set the width and height of the Gaussian profile as free parameters. We conducted Markov chain Monte Carlo (MCMC) simulations with the \texttt{emcee} tool \citep{Mackey2013} to sample from the posterior. The noise of each CCF was assigned as its standard deviation. The MCMC calculation was performed on the whole CCF matrix without the secondary eclipse data. The best-fit CCF-map is presented in the middle panel of Fig.~\ref{CCF-map}. 
The MCMC yields estimates of $176.7\pm0.6$ km\,s$^{-1}$ for $K_\mathrm{p}$ and $1.0\pm0.2$ km\,s$^{-1}$ for $\mathrm{\Delta} \varv$.  The small $\mathrm{\Delta} \varv$ could be a signature of atmospheric dynamics, but it could also originate from the uncertainties of the stellar systemic RV and planetary orbital ephemeris, which typically produce a change of $\mathrm{\Delta} \varv$ of several km\,s$^{-1}$ \citep{Yan2020}.
The bottom panel of Fig.\ref{CCF-map} presents the CCFs shifted to the planetary rest frame using the best-fit $K_\mathrm{p}$ value.


The best-fit $K_\mathrm{p}$ is consistent with the theoretical values within $2 \sigma$ (cf. Table \ref{paras_K20}). These values were calculated using the equation
\begin{equation}
 K_\mathrm{p} = (2 \mathrm{\pi} G \cdot M_\star / P)^{1/3} \cdot \mathrm{sin}\,i_\mathrm{p}     ,
\end{equation}
  where $G$ is the gravitational constant, $P$ is the orbital period, $M_\star$ is the stellar mass, and $i_\mathrm{p}$ is the orbital inclination. 
On the other hand, the well-determined $K_\mathrm{p}$ value from planetary emission spectroscopy can be used to calculate the mass of the star. Using the above equation, we obtained $M_\star = 2.00\pm0.02 M_\odot$, which is slightly higher than the literature values inferred from theoretical stellar evolution models: $1.89_{-0.05}^{+0.06} M_\odot$ \citep{Talens2018} and $1.76_{-0.14}^{+0.19} M_\odot$ \citep{Lund2017}. This demonstrates that emission spectroscopy is a unique tool for independently measuring stellar mass. This was initially proposed and demonstrated by \cite{deKoK2013}.

\begin{figure}
   \centering
   \includegraphics[width=0.5\textwidth]{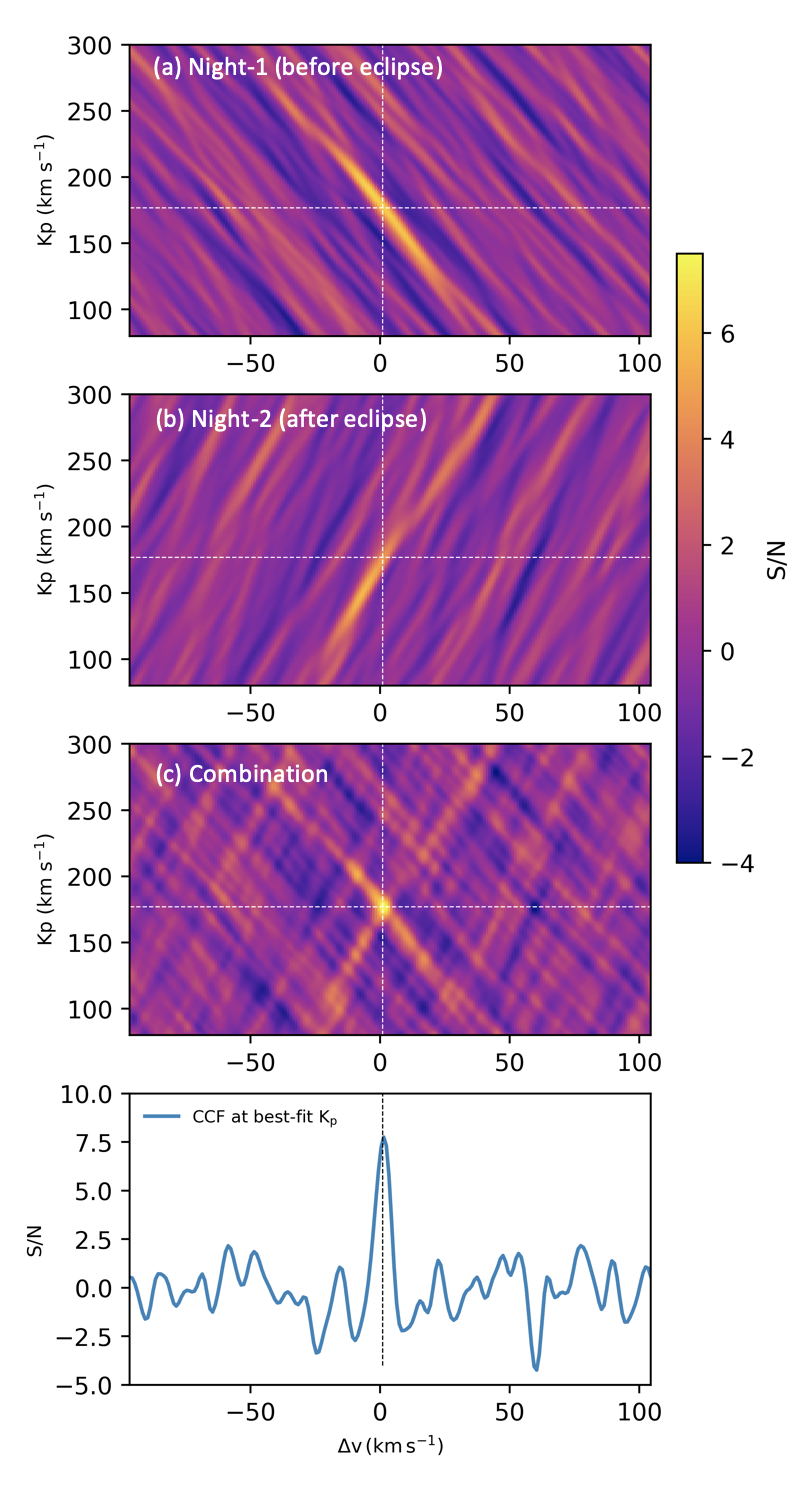}
      \caption{Combined cross-correlation functions with different $K_\mathrm{p}$ values (i.e., $K_\mathrm{p}$-$\mathrm{\Delta} \varv$ map). \textit{Panels a, b, and c:} Maps of night 1, night 2, and the combination of the two nights, respectively. The dashed white lines indicate the location of the best-fit $K_\mathrm{p}$ and $\mathrm{\Delta} \varv$ values (i.e., 177.5 km\,s$^{-1}$ and 1.0 km\,s$^{-1}$).
      \textit{Panel d:} CCF at the best-fit $K_\mathrm{p}$.
      }
         \label{Kp-map}
   \end{figure}

We also computed the classical $K_\mathrm{p}$-$\mathrm{\Delta} \varv$ maps for the individual nights and the combined data (Fig.\ref{Kp-map}). The maps were generated by adding up the CCFs in the planetary rest frame with different $K_\mathrm{p}$ values. To estimate the significance of the detection, we computed the standard deviation of the CCFs within the $\left| \mathrm{\Delta} \varv \right|$ range between 100 and 200 km\,s$^{-1}$, and took this value as the noise of the map. The detection significance of night 1 is higher than that of night 2. The $K_\mathrm{p}$-$\mathrm{\Delta} \varv$ map of the combined data shows a clear peak (S/N $\sim$ 7.7) located around $K_\mathrm{p}$ = $177$ km\,s$^{-1}$ and $\mathrm{\Delta} \varv$ = $1.0$ km\,s$^{-1}$.

The detection of \ion{Fe}{i} emission lines in the thermal emission spectrum of KELT-20b is  unambiguous evidence for a thermal inversion layer in its dayside atmosphere. 
KELT-20b is the fourth planet in which \ion{Fe}{i} emission lines are detected on the planetary dayside hemisphere, after KELT-9b, WASP-189b, and WASP-33b \citep{Pino2020, Yan2020, Nugroho2020W33,Cont2021}. These four planets are all UHJs orbiting A-type stars, indicating that hot stars can create strong temperature inversion, which makes the \ion{Fe}{i} emission lines relatively strong. This scenario is consistent with the simulation by \cite{Lothringer2019}, who proposed that the temperature and slope of the inversion layer increase with stellar effective temperature because of the enhanced absorption at short wavelengths and low pressures.

In addition to \ion{Fe}{i}, we also searched for other species, including \ion{Fe}{ii}, \ion{Ti}{i}, \ion{Ti}{ii}, TiO, VO, and FeH. However, we were not able to detect them (cf. Fig.~\ref{APP-non-detection}). The nondetection of these species could be due to several reasons, for example, relatively weak emission lines in the CARMENES wavelength range, a poor accuracy of the line lists, an insufficient S/N of the data, or the nonexistence of the species in the temperature inversion layer. The detection of the strong \ion{Fe}{i} feature benefits from the high-accuracy line list and the large number of spectral lines covering a wide wavelength range.

%

\subsubsection{Differences between the results of the two nights}
We observed the thermal emission spectrum during two nights, before eclipse for the first night and after eclipse for the second night (Fig.\ref{fig-obs}). The cross-correlation results (i.e., CCF and $K_\mathrm{p}$ map) of each night are presented in Fig.\ref{CCF-map} and Fig.\ref{Kp-map}.
There are some differences between the $K_\mathrm{p}$ maps of the two nights. 
The maximum S/N (6.4 $\sigma$) for night 1 is located at $K_\mathrm{p}$ = $176.4\pm4$ km\,s$^{-1}$ and $\mathrm{\Delta} \varv$ = $1.2_{-1.6}^{+1.7}$ km\,s$^{-1}$, which is consistent with the expected $K_\mathrm{p}$. However, the maximum S/N (5.7 $\sigma$) for night 2 is located at $K_\mathrm{p}$ = $152.4\pm{3.2}$ km\,s$^{-1}$ and $\mathrm{\Delta} \varv$ = $-9.5\pm{0.9}$ km\,s$^{-1}$, deviating from the expected values. This difference means that the RVs of the detected \ion{Fe}{i} emission lines deviate by several km\,s$^{-1}$ from the orbital motion. This deviation might be the result of atmospheric dynamics such as rotation and winds. In addition, the signal from the before-eclipse observation is slightly stronger than that from the after-eclipse observation. This asymmetry feature might be due to an eastward hotspot offset, which causes the average temperature of the visible hemisphere during the before-eclipse observation to be higher than that during the after-eclipse observation. However, the poor S/N of the second night is not sufficient to draw any conclusions about the nature of dynamics in the atmosphere. Further observations with higher S/N in combination with general circulation models should be able to confirm or discard this hypothesis.

In addition to the emission spectroscopy, \ion{Fe}{i} has also been detected in the transmission spectrum of KELT-20b. The \ion{Fe}{i} transmission spectrum is blueshifted by 5 - 10 km\,s$^{-1}$ and the signal has an asymmetric feature between the first and second half of the transit \citep{Stangret2020, Nugroho2020-KELT20, Hoeijmakers2020, Rainer2021}, probably caused by atmospheric dynamics.

\subsection{Retrieval of atmospheric properties}
\subsubsection{Retrieval with CARMENES data}
Techniques to retrieve the properties of exoplanet atmospheres from high-resolution spectroscopy have been developing in recent years \citep[e.g.,][]{Brogi2019, Gandhi2019,Shulyak2019, Gibson2020}.
To retrieve the atmospheric properties of KELT-20b from the observed \ion{Fe}{i} emission lines, we applied the method described in \cite{Yan2020} with the following steps:\\
\textit{(1) Calculating a master residual spectrum.} Because $K_\mathrm{p}$ and $\mathrm{\Delta} \varv$ are well determined with our data, we fixed these two parameters and shifted all the residual spectra by correcting the planetary orbital RV and $\mathrm{\Delta} \varv$. Only the out-of-eclipse data were used. The master residual spectrum was then obtained by averaging these shifted residual spectra with the square of the S/N as the weight of each data point. This master residual spectrum is regarded as the normalized spectrum with no information about the spectral continuum, because we have already normalized the original spectra and removed the broad features when we performed the \texttt{SYSREM} and Gaussian filtering described in Section 2. \\
\textit{(2) Setting up the spectral model.} We used the \texttt{petitRADTRANS} tool to forward-model the dayside spectrum. The $T$-$P$ profile was parameterized using the two-point model. The atmosphere model consisted of 25 layers that were uniformly spaced in log($P$) ranging from 1 bar to $10^{-8}$ bar. We used an opacity grid of \ion{Fe}{i} up to 25\,000\,K.
For a given $T$-$P$ profile, the mixing ratio of \ion{Fe}{i} was computed with the chemical equilibrium module \texttt{easyCHEM} of \texttt{petitCode} \citep{Molliere2015, Molliere2017}. We also set the elemental abundance to solar and log\,$g$ to 3.0. 
The simulated thermal emission spectrum of \ion{Fe}{i} (1 + $F_\text{p}$/$F_\text{s}$) was normalized, convolved, interpolated, and filtered in the same way as described in Section 3.1. \\
\textit{(3) Fitting the master residual spectrum with the spectral model.} Following  \cite{Yan2020}, we assumed a standard Gaussian likelihood function expressed in logarithm,
\begin{equation}
      \mathrm{ln}(L) = -\frac{1}{2}\sum_{i} \left[ \frac{(R_i - m_i)^2}{(\beta \sigma_i)^2} + \mathrm{ln}(2 \pi (\beta \sigma_i)^2) \right] ,
\label{eq-CAR}
\end{equation}
where $R_i$ is the observed master residual spectrum at wavelength point $i$, $m_i$ is the spectral model, $\sigma_i$ is the uncertainty of the observed spectrum, and $\beta$ is a uniform scaling term of the uncertainty. We then applied the MCMC method to obtain the best-fit parameters and their uncertainties by evaluating the likelihood function with the \texttt{emcee} tool. The MCMC calculation had 5000 steps with 24 walkers. We set uniform priors for the parameters with the boundary conditions shown in Table \ref{tab-mcmc}.

\begin{figure}
   \centering
   \includegraphics[width=0.5\textwidth]{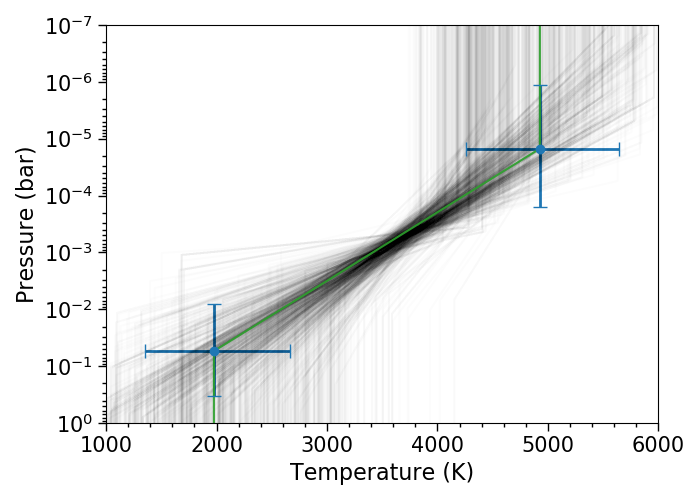}
   \includegraphics[width=0.5\textwidth]{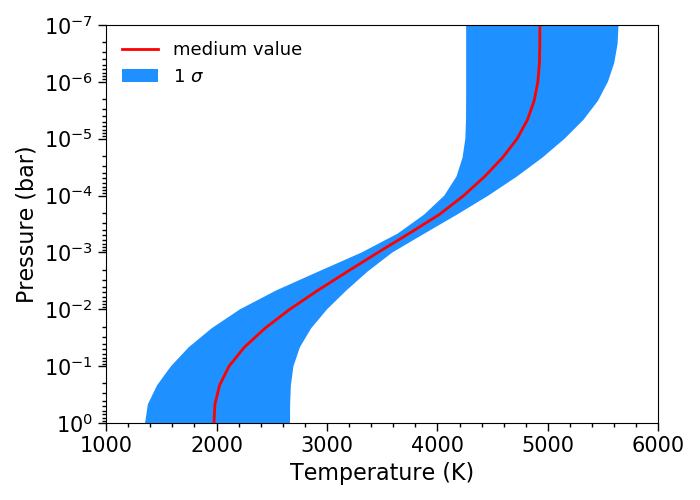}
      \caption{Retrieved $T$-$P$ profile from the CARMENES data. \textit{Upper panel:} Retrieved results of the two-point $T$-$P$ model. The blue points with error bars are the best-fit ($T_\mathrm{1}$, $P_\mathrm{1}$) and ($T_\mathrm{2}$, $P_\mathrm{2}$) values. The gray lines show examples of the $T$-$P$ profiles sampled by the MCMC analysis.
      \textit{Lower panel:} Median of the sampled $T$-$P$ profiles (red line) and the 1$\sigma$ envelope (blue shadow). They are generated by sorting the temperatures of the MCMC samples at each of the 25 atmosphere layers.
      }
         \label{TP-mcmc}
   \end{figure}

The retrieved parameters are summarized in Table \ref{tab-mcmc} and the best-fit $T$-$P$ profile is presented in Fig.\ref{TP-mcmc}. The posterior distributions are plotted in Fig.\ref{APP-corner}. The retrieved result indicates that the planet has a steep temperature inversion with very high temperatures ($\sim$ 4900\,K) in the upper layer. 

In this retrieval, we assumed a solar elemental abundance for Fe and set log\,$g$ to 3.0. However, the retrieved $T$-$P$ profile depends on the Fe elemental abundance and on surface gravity. We tested different abundances and surface gravity values and found that these two parameters mostly affect the location (i.e., $P_\mathrm{1}$ and $P_\mathrm{2}$) of the inversion, while the temperature is less affected. With a higher elemental abundance or a lower surface gravity, the retrieved inversion layer is located at higher altitudes.
We performed the retrieval with log\,$g$ as a free parameter with an upper boundary of 3.4 and lower boundary of 2.5. The retrieved results indicate that the $T$-$P$ profile is heavily degenerate with log\,$g$ (Appendix Fig.\ref{APP-logg}) and the data constrain  log\,$g$ only poorly. Future RV follow-up observations will be useful in constraining the planetary mass and the planetary surface gravity.

We also performed the retrieval with the Fe elemental abundance ([Fe/H]) as a free parameter. For a given [Fe/H] value, we calculated the mixing ratio of \ion{Fe}{i} using \texttt{easyCHEM}.
The retrieved [Fe/H] is $0.7_{-1.4}^{+1.3}$ dex (Table \ref{TP-mcmc} and Fig.\ref{APP-FeH}). The large error of the retrieved value indicates that there is a certain degree of degeneracy between [Fe/H] and the $T$-$P$ profile. 
In addition, \cite{Fossati2021-NLTE} found that
nonlocal thermodynamic equilibrium (NLTE) can affect the upper atmosphere of UHJs. NLTE effects at low pressures can alter the Fe level population and thereby affect the retrieval of the Fe abundance and $T$-$P$ profile.
Detection of other species (e.g., \ion{Fe}{ii} and FeH) and inclusion of the NLTE effects in the retrieval will enable us to better constrain the Fe abundance.

%
\begin{table*}
\caption{Best-fit parameters from the $T$-$P$ profile retrieval.}             
\label{tab-mcmc}      
\centering                          
\begin{threeparttable}
        \begin{tabular}{l c c c c c c}        
        \hline\hline \noalign{\smallskip}                 
                Parameter & Value & Value (TESS included)  &  Value ([Fe/H] free) & Value (log\,$g$ free)  & Boundaries & Unit \\     
        \hline     \noalign{\smallskip}                   
  \rule{0pt}{2.5ex} $T_\mathrm{1}$ & $4900\pm{700}$ & $4900_{-600}^{+700}$ & $4700_{-600}^{+800}$ & $4900\pm{700}$ & 1000 to 6000 & K\\ 
  \rule{0pt}{2.5ex} log $P_\mathrm{1}$ & $-4.8_{-1.1}^{+1.0}$ & $-5.0_{-1.1}^{+1.0}$ & $-5.5_{-1.0}^{+1.5}$ & $-4.9\pm{1.0}$ & $-7$ to 0 & log bar \\    
  \rule{0pt}{2.5ex} $T_\mathrm{2}$ & $1900_{-600}^{+700}$ & $2550_{-250}^{+150}$ & $2000\pm{700}$ & $1900_{-600}^{+700}$ & 1000 to 6000 & K \\ 
  \rule{0pt}{2.5ex} log $P_\mathrm{2}$ & $-1.3\pm{0.8}$ & $-1.5_{-0.6}^{+0.7}$ & $-1.7_{-1.5}^{+1.1}$ & $-1.2_{-0.8}^{+0.7}$ & $-7$ to 0 & log bar \\  
  \rule{0pt}{2.5ex} $\beta$ & $0.752\pm{0.0015}$ & $0.752\pm{0.0015}$ & $0.752\pm{0.0015}$ & $0.752\pm{0.0015}$ & 0 to 10 & ...\\     
  \rule{0pt}{2.5ex} [Fe/H] & 0 (fixed) & 0 (fixed) & $0.7_{-1.4}^{+1.3}$ & 0 (fixed) & $-3$ to +3 & dex \\  
  \rule{0pt}{2.5ex} log\,$g$ & 3.0 (fixed) & 3.0 (fixed) & 3.0 (fixed) & $3.0\pm0.3$ & 2.5 to 3.4 & log in cgs \\              
\noalign{\smallskip}  \hline                                   
        \end{tabular}
\end{threeparttable}      
\end{table*}

\subsubsection{Joint retrieval with TESS data}
The secondary eclipse of KELT-20b was recently measured by \cite{Wong2021} using the TESS light curve. The reported eclipse depth is $111_{-36}^{+35}$\,ppm. This provides the spectral continuum level of the dayside hemisphere. Therefore, the TESS data are complementary to the high-resolution spectrum of \ion{the Fe}{i} emission lines, which lacks the continuum information. We therefore included the TESS eclipse data point in our retrieval. 
First, we calculated the un-normalized $F_\text{p}$/$F_\text{s}$ spectrum, which contains both the continuum spectrum and the line spectrum. Then we integrated the flux spectrum using the \texttt{rebin-give-width} tool of \texttt{petitRADTRANS} from 0.6\,$\mathrm{\mu}$m to 1.0\,$\mathrm{\mu}$m, which is an approximation of the TESS bandpass. In this way, we obtained the modeled secondary eclipse depth. Subsequently, we calculated the likelihood function for the TESS data as
\begin{equation}
      \mathrm{ln}(L_\mathrm{T}) = -\frac{1}{2} \left[ \frac{(R_\mathrm{T} - m_{T})^2}{\sigma_\mathrm{T}^2} + \mathrm{ln}(2 \pi \sigma_\mathrm{T}^2) \right],
\label{eq-TESS}
\end{equation}
where $R_\mathrm{T}$ is the modeled eclipse depth, $m_\mathrm{T}$ is the TESS measured eclipse depth (111\,ppm), and $\sigma_\mathrm{T}$ is the noise (36\,ppm). The subscript (T) denotes that these are parameters for the TESS calculation.
This likelihood function was then added to the likelihood function of the CARMENES data (Eq.~\ref{eq-CAR}). To perform the joint retrieval, we applied the MCMC calculations to this combined likelihood function in the same way as for the CARMENES-only retrieval.

The jointly retrieved values are presented in Table \ref{tab-mcmc} and the $T$-$P$ profile envelope is plotted in Fig. \ref{TP-HELIOS}. The TESS data mostly constrain the cooler low-altitude layers of the atmosphere in which the continuum spectrum originates, therefore, the $T_\mathrm{2}$ value ($2550_{-250}^{+150}$\,K) is better determined than the result of CARMENES-only data (cf. the posterior distribution plot in Fig.\ref{APP-corner-TESS}).

In the retrieval, we only included the blackbody thermal emission as the continuum source. However, $\mathrm{H^{-}}$ has been proposed as an important continuum source for UHJs \citep[e.g.,][]{Parmentier2018}. Therefore, we estimated the impact of $\mathrm{H^{-}}$ on the retrieval. We took the best-fit parameters from the retrieval and compared the modeled spectra with and without $\mathrm{H^{-}}$.
For the TESS secondary eclipse of KELT-20b, the $\mathrm{H^{-}}$ contribution is $\sim$ 10\,ppm, which is within the noise level of the measured eclipse depth. For the strength of the Fe emission line, $\mathrm{H^{-}}$ has a negligible contribution \citep{Yan2020}.

We assumed that the measured dayside flux arises from the thermal emission. However, reflected light can also contribute to the measured TESS eclipse depth \citep[e.g.,][]{Daylan2021,vonEssen2021}. Because of the degeneracy between the reflected flux and the thermal emission flux, the TESS data alone cannot constrain the geometric albedo ($A_\mathrm{g}$) of KELT-20b \citep{Wong2021}. Nevertheless, measurements of other UHJs suggest that their albedos are very low \citep[e.g.,][]{Bell2017, Shporer2019, Wong2021}.
To estimate the strength of possible reflected light, we followed the method described by \cite{Alonso2018} and calculated the light contribution to $F_\text{p}$/$F_\text{s}$ as
$A_\mathrm{g}$\,$R_\mathrm{p}^2/a^2$. When assuming $A_\mathrm{g}$ = 0.1, we obtained a value of 24\,ppm, which is within the noise level of the TESS result.


\subsubsection{Comparison with self-consistent models}
To compare the retrieved $T$-$P$ profile with self-consistent models, we calculated $T$-$P$ profiles using the \texttt{HELIOS} code originally presented in \cite{Malik2017}. 
We used an updated version of \texttt{HELIOS}, which includes opacities due to neutral and singly ionized species as described in \cite{Fossati2021-NLTE}. In particular, we included atomic line opacities due to neutral and singly ionized
atoms, namely \ion{C}{i-ii}, \ion{Cr}{i-ii}, \ion{Fe}{i-ii}, \ion{K}{i-ii}, \ion{Mg}{i-ii}, \ion{Na}{i-ii}, \ion{O}{i-ii}, and \ion{Si}{i-ii}, which are found to contribute most
to the line opacity throughout the planetary atmosphere. The
original line lists are those produced by R.~Kurucz\footnote{\url{http://kurucz.harvard.edu/linelists/gfall/}} \citep{Kurucz2018}.
The molecular line opacity includes molecules such as CH$_{\rm 4}$,
CO$_{\rm 2}$, CO, H$_{\rm 2}$O, HCN, NH$_{\rm 3}$, OH, SiO, TiO, and VO. The pretabulated cross sections of these molecules were taken from the
public opacity database for exoplanetary atmospheres\footnote{\url{https://dace.unige.ch/opacity}}. We also extended continuum opacity sources by including, for example, continuum transitions
of H$^-$, He$^-$, and metals \citep[see][]{Fossati2020, Fossati2021-NLTE}. The \texttt{HELIOS} calculations were performed assuming local thermodynamic equilibrium.

\begin{figure}
   \centering
   \includegraphics[width=0.5\textwidth]{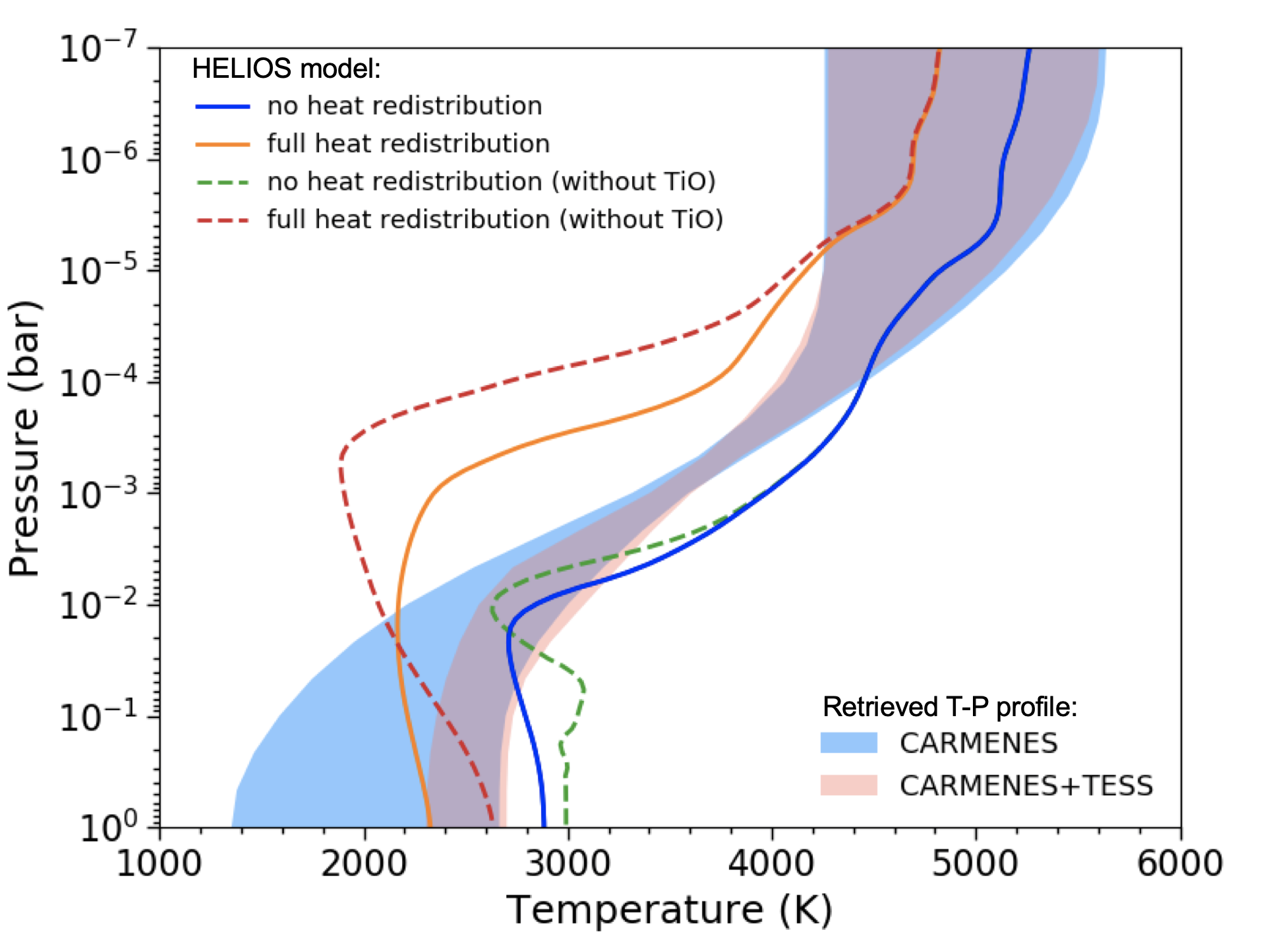}
      \caption{Comparison between the theoretical $T$-$P$ profile from the self-consistent \texttt{HELIOS} model and the retrieved $T$-$P$ profile from the CARMENES and TESS observations. The solid lines are the \texttt{HELIOS} results assuming no heat redistribution or full heat redistribution from the dayside to nightside. The dashed lines are the models without TiO. The 1$\sigma$ range of the retrieved $T$-$P$ profiles is indicated as the blue shadow (CARMENES data) and the salmon shadow (CARMENES + TESS data).
      }
         \label{TP-HELIOS}
   \end{figure}

We assumed solar abundance, log\,$g$ = 3.0, and zero albedo to compute the atmosphere of KELT-20b. We calculated two extreme cases: one case with full heat redistribution from the dayside to nightside (corresponding to a dayside $T_\mathrm{eq}$ of 2300\,K), and the other case without heat redistribution (corresponding to dayside $T_\mathrm{eq} \sim 3000\,K$). In addition, we calculated the models with an atmosphere without TiO. The modeled $T$-$P$ profiles are presented in Fig. \ref{TP-HELIOS}. 

All of these self-consistent models predict the existence of a temperature inversion layer. The contribution of TiO to the temperature inversion is relatively weak, especially for the $T_\mathrm{eq}$ $\sim$ 3000\,K case. The retrieved temperature inversion layer is located between the two heat redistribution cases, especially at pressures of $10^{-3}-10^{-4}$ bar, where the Fe emission line cores are formed. This means that our retrieved $T$-$P$ profile is consistent with the self-consistent models because the actual heat redistribution is expected to be in between the two extreme cases.

%

%
%
%
%
%

\section{Conclusions}
We observed the dayside thermal emission spectrum of the ultra-hot Jupiter KELT-20b with the CARMENES spectrograph. The observation covers planetary orbital phases before and after the secondary eclipse. We employed the cross-correlation technique to search for atmospheric species in the planetary dayside hemisphere and detected a strong neutral Fe signal. The detected Fe lines are in emission, which unambiguously indicates the existence of a temperature inversion layer in the atmosphere. So far, temperature inversion has been detected in four UHJs (i.e., KELT-9b, WASP-33b, WASP-189b, and KELT-20b) using high-resolution thermal emission spectroscopy.
The detection of temperature inversion is consistent with theoretical simulations that predict its existence.

We retrieved the atmospheric profile with the observed high-resolution \ion{Fe}{i} emission lines in KELT-20b using the \texttt{petitRADTRANS} forward model and the \texttt{easyCHEM} chemical equilibrium code.
The results show a strong temperature inversion with a temperature around 4900\,K at the upper layer of the inversion.  In addition, we included the secondary eclipse depth that was
recently measured with TESS \citep{Wong2021}. The joint CARMENES + TESS fit yields a tighter constraint on the temperature ($\sim$ 2550\,K) of the lower-altitude atmosphere in which the photosphere is located.
We also computed the self-consistent atmospheric structure using the code \texttt{HELIOS,}  which shows $T$-$P$ profiles that are consistent with the retrieved result.

Ground-based high-resolution emission spectroscopy is a powerful technique for probing the dayside hemispheres of UHJs. Emission spectroscopy is particularly sensitive in characterizing the temperature structure (e.g., temperature inversion). In addition, the phase-resolved emission spectroscopy can be used to characterize atmospheric dynamics and the global distribution of temperature and chemical species.


\begin{acknowledgements}
We thank the referee for the useful comments.
F.Y. acknowledges the support of the the DFG Research Unit FOR2544 ``Blue Planets around Red Stars'' (RE 1664/21-1).
CARMENES is an instrument for the Centro Astron\'omico Hispano-Alem\'an (CAHA) at Calar Alto (Almer\'{\i}a, Spain), operated jointly by the Junta de Andaluc\'ia and the Instituto de Astrof\'isica de Andaluc\'ia (CSIC).
  
  CARMENES was funded by the Max-Planck-Gesellschaft (MPG), 
  the Consejo Superior de Investigaciones Cient\'{\i}ficas (CSIC),
  the Ministerio de Econom\'ia y Competitividad (MINECO) and the European Regional Development Fund (ERDF) through projects FICTS-2011-02, ICTS-2017-07-CAHA-4, and CAHA16-CE-3978, 
  and the members of the CARMENES Consortium 
  (Max-Planck-Institut f\"ur Astronomie,
  Instituto de Astrof\'{\i}sica de Andaluc\'{\i}a,
  Landessternwarte K\"onigstuhl,
  Institut de Ci\`encies de l'Espai,
  Institut f\"ur Astrophysik G\"ottingen,
  Universidad Complutense de Madrid,
  Th\"uringer Landessternwarte Tautenburg,
  Instituto de Astrof\'{\i}sica de Canarias,
  Hamburger Sternwarte,
  Centro de Astrobiolog\'{\i}a and
  Centro Astron\'omico Hispano-Alem\'an), 
  with additional contributions by the MINECO, 
  the Deutsche Forschungsgemeinschaft through the Major Research Instrumentation Programme and Research Unit FOR2544 ``Blue Planets around Red Stars'', 
  the Klaus Tschira Stiftung, 
  the states of Baden-W\"urttemberg and Niedersachsen, 
  and by the Junta de Andaluc\'{\i}a.
  
  Based on data from the CARMENES data archive at CAB (CSIC-INTA).
  
  We acknowledge financial support from the Agencia Estatal de Investigaci\'on of the Ministerio de Ciencia, Innovaci\'on y Universidades and the ERDF  through projects PID2019-109522GB-C51/2/3/4, PGC2018-098153-B-C33, AYA2016-79425-C3-1/2/3-P, ESP2016-80435-C2-1-R and the Centre of Excellence ``Severo Ochoa'' and ``Mar\'ia de Maeztu'' awards to the Instituto de Astrof\'isica de Canarias (SEV-2015-0548), Instituto de Astrof\'isica de Andaluc\'ia (SEV-2017-0709), and Centro de Astrobiolog\'ia (MDM-2017-0737), and the Generalitat de Catalunya/CERCA programme.  
  
  T.H. and P.M. acknowledge support from the European Research Council under the Horizon 2020 Framework Program via the ERC Advanced Grant Origins 83 24 28.
   N.C. and A.S.L. acknowledge funding from the European Research Council under the European Union's Horizon 2020 research and innovation program under grant agreement No 694513.

\end{acknowledgements}

\bibliographystyle{aa} 

\bibliography{K20-Fe-refer}

\begin{appendix}
\section{Additional figures}

   \begin{figure*}
   \centering
   \includegraphics[width=0.98\textwidth]{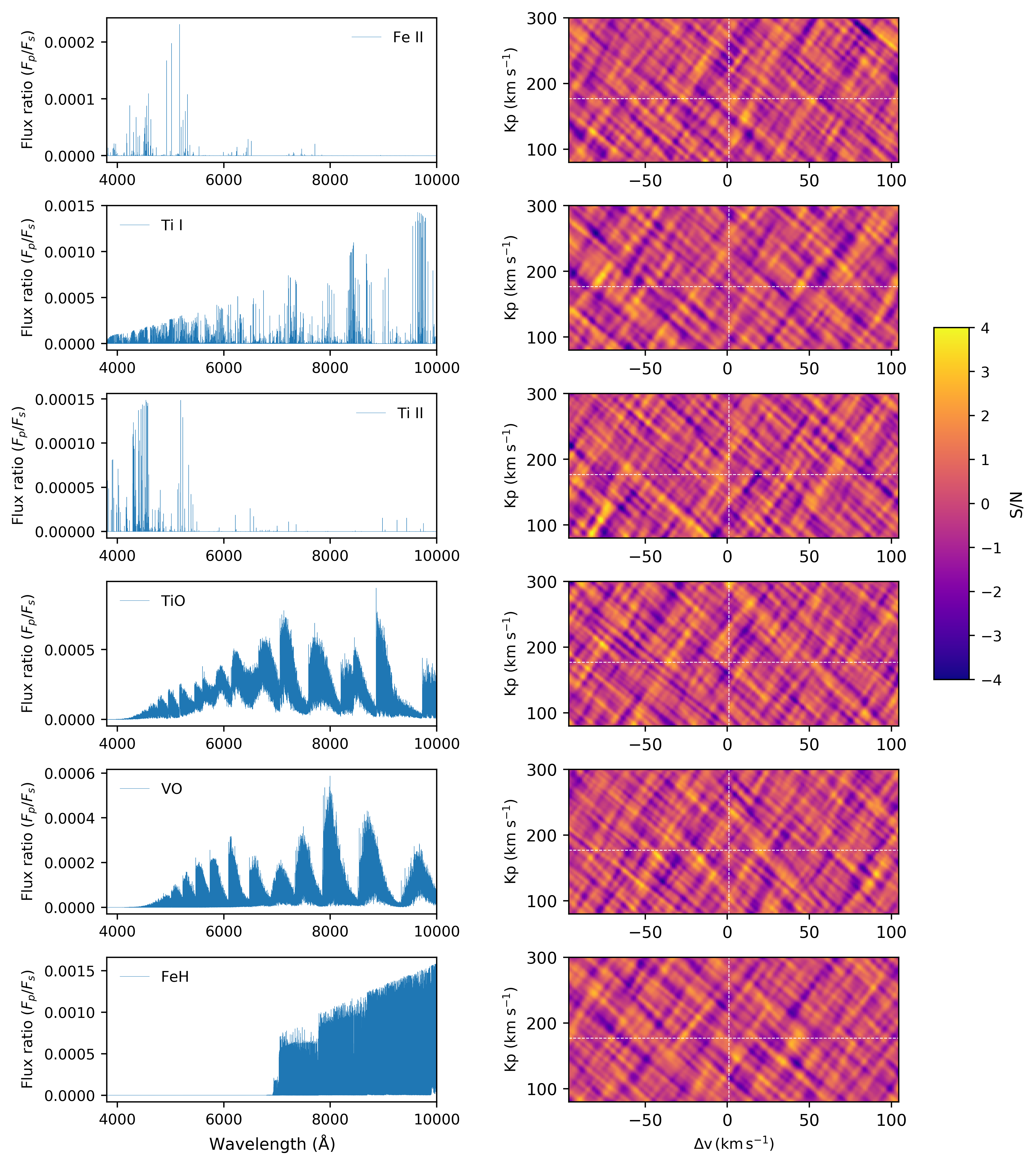}
      \caption{Nondetection of several other chemical species. \textit{Left panels:} Spectral model of each species. These are normalized spectra that are calculated in a similar way as described in Section 3.1.1. \textit{Right panels:} Combined two-night $K_\mathrm{p}$-$\mathrm{\Delta} \varv$ map of each species. The dashed white lines indicate the location of the best-fit $K_\mathrm{p}$-$\mathrm{\Delta} \varv$ from the \ion{Fe}{i} signal. 
      }
         \label{APP-non-detection}
   \end{figure*}

   \begin{figure*}
   \centering
   \includegraphics[width=0.6\textwidth]{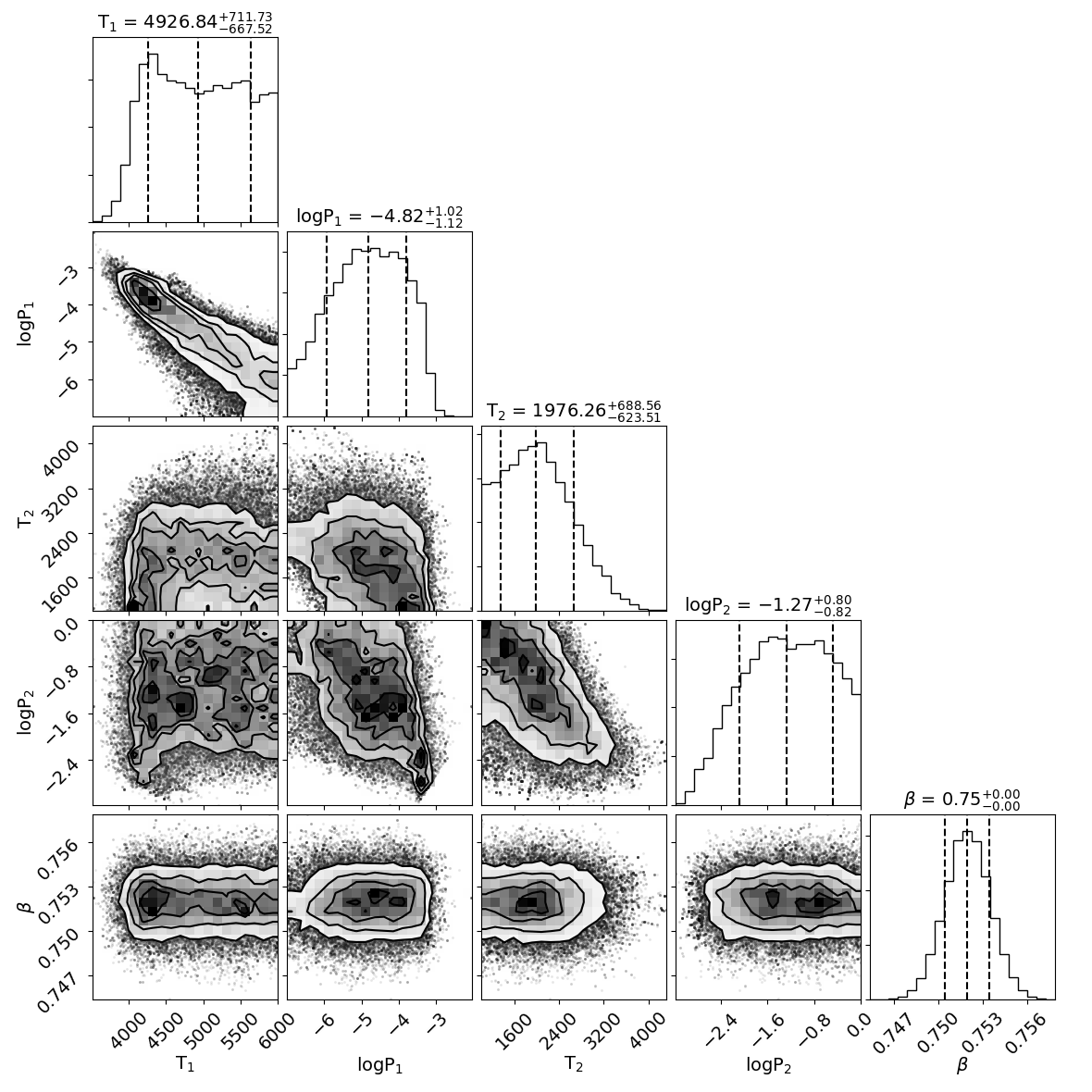}
      \caption{Posterior distribution of the parameters from the MCMC fit of the CARMENES data. Here both $\mathrm{log}\,g$ and [Fe/H] are fixed.}
         \label{APP-corner}
   \end{figure*}

   \begin{figure*}
   \centering
   \includegraphics[width=0.65\textwidth]{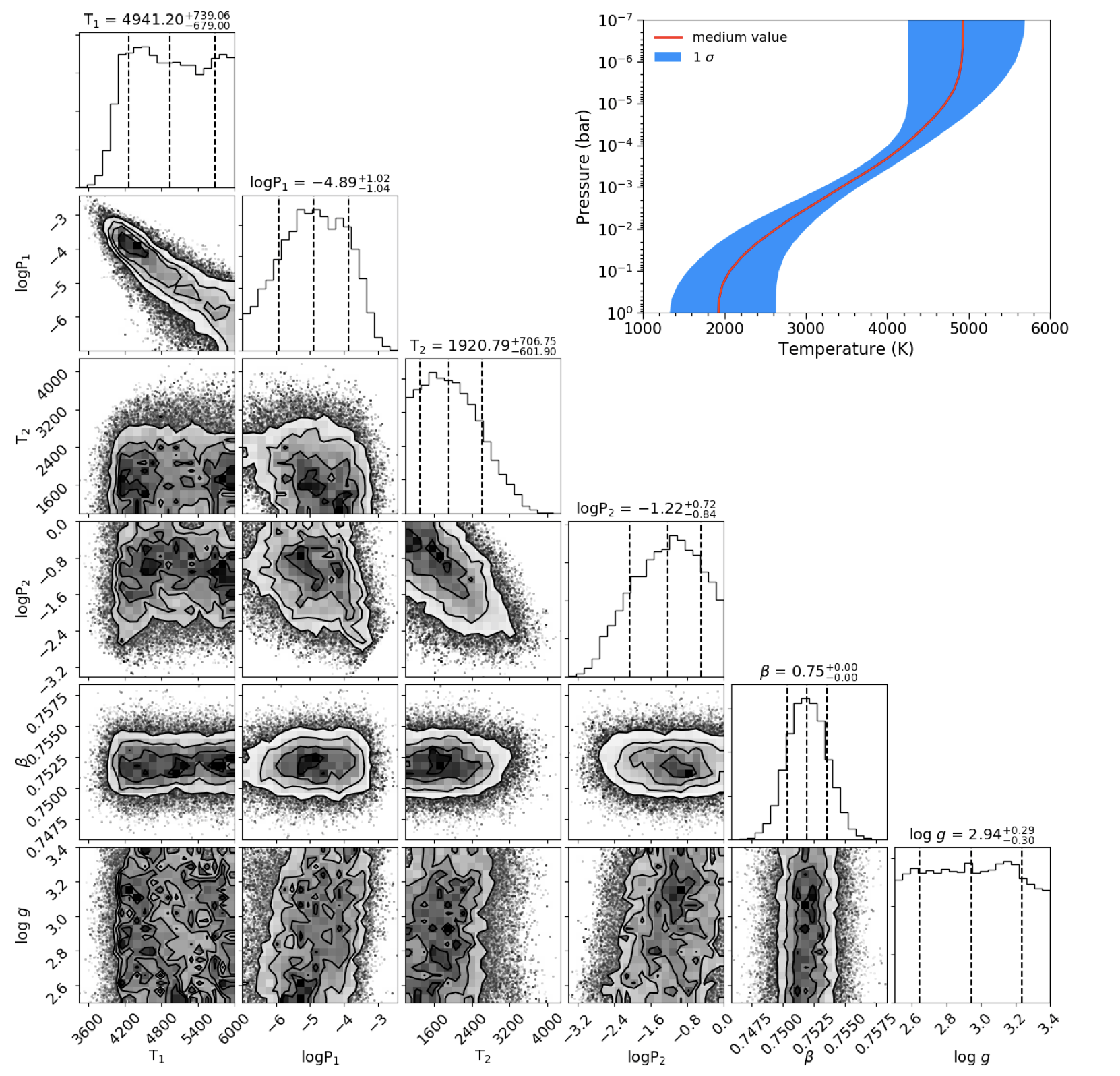}
      \caption{Same as Fig.\ref{APP-corner}, but with $\mathrm{log}\,g$ as a free parameter. The inset shows the retrieved $T$-$P$ profile.}
         \label{APP-logg}
   \end{figure*}

   \begin{figure*}
   \centering
   \includegraphics[width=0.65\textwidth]{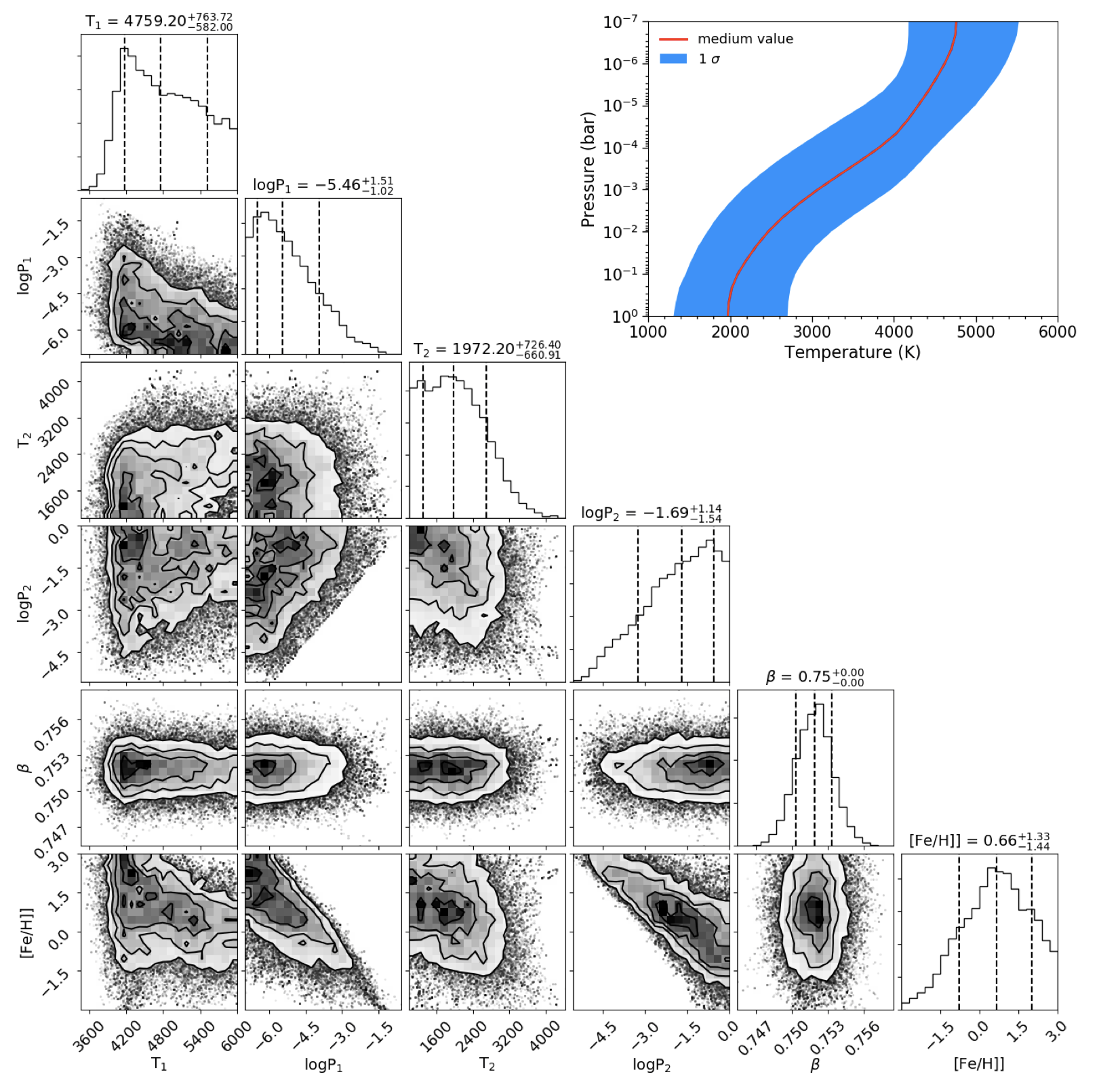}
      \caption{Same as Fig.\ref{APP-corner}, but with [Fe/H] as a free parameter. The inset shows the retrieved $T$-$P$ profile.}
         \label{APP-FeH}
   \end{figure*}

   \begin{figure*}
   \centering
   \includegraphics[width=0.6\textwidth]{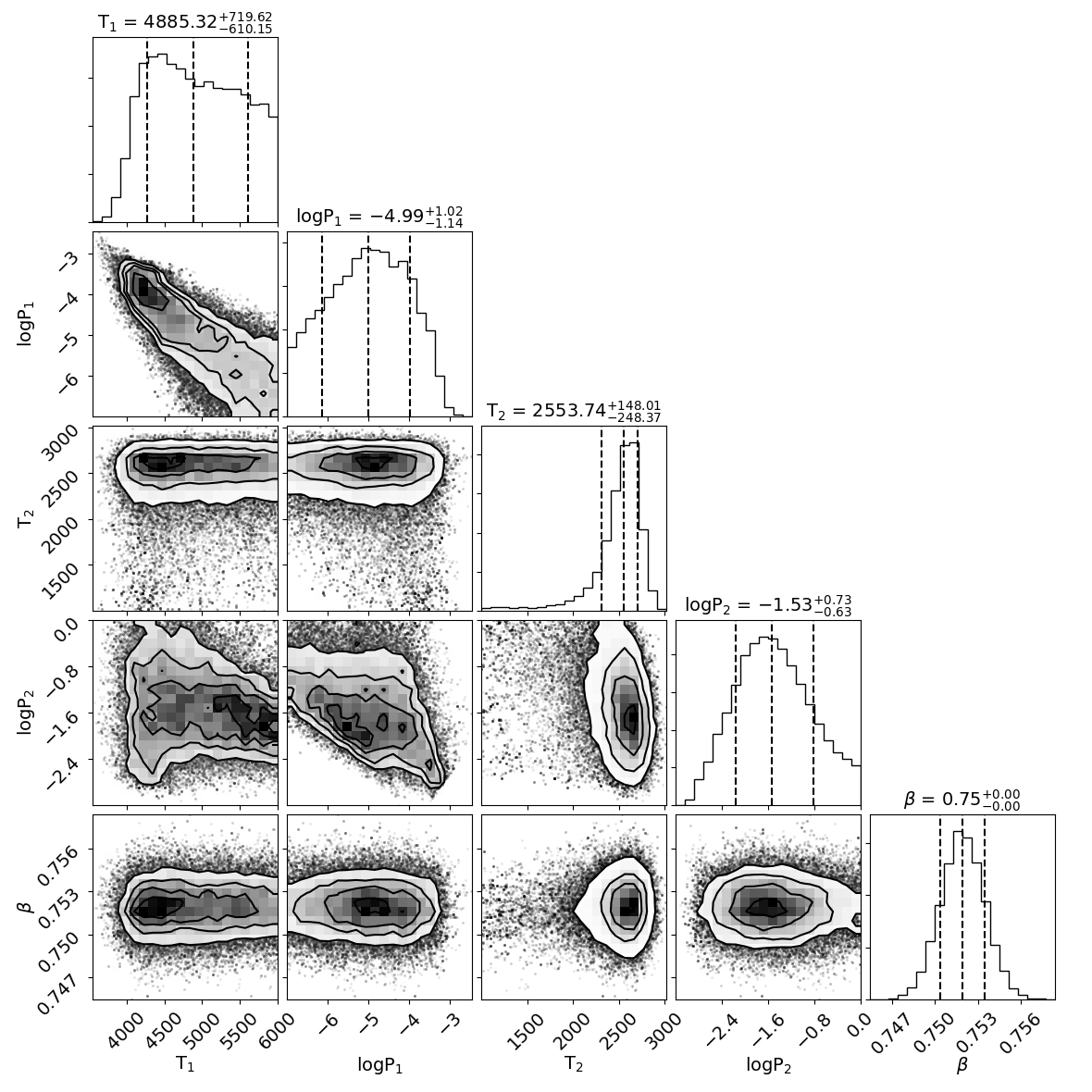}
      \caption{Same as Fig.\ref{APP-corner}, but for the CARMENES+TESS data.}
         \label{APP-corner-TESS}
   \end{figure*}

\end{appendix}

\end{document}